\documentclass[aps,prb,twocolumn]{revtex4}

\usepackage{amsmath}

\usepackage[latin1]{inputenc}

\usepackage{color}

\usepackage[ps2pdf,
bookmarks=false,
pdfstartview=FitH,
pdfpagelayout=OneColumn,
pdfborder={0 0 5},
citebordercolor={0 0 1},
linkbordercolor={0 0 1},
urlbordercolor={0 0 1}
]{hyperref}

\usepackage[sort&compress]{natbib}

\usepackage[figure]{hypcap}

\usepackage{graphicx}

\begin{document}

\title{Domain-wall profile in the presence of anisotropic exchange interactions:\\ Effective on-site anisotropy}

\author{A. O. García Rodríguez}
\email{alexis@ifi.unicamp.br}

\author{A. Villares Ferrer}

\author{A. O. Caldeira}

\affiliation{Instituto de Física Gleb Wataghin, Departamento de Física da Matéria Condensada, Universidade Estadual de Campinas, 13083-970, Campinas, São Paulo, Brazil}

\begin{abstract}
Starting from a $D$-dimensional $XXZ$ ferromagnetic Heisenberg model in an hypercubic lattice, it is demonstrated that the anisotropy in the exchange coupling constant leads to a $D$-dependent effective on-site anisotropy interaction often ignored for $D>1$. As a result the effective width of the wall depends on the dimensionality of the system. It is shown that the effective one-dimensional Hamiltonian is not the one-dimensional $XXZ$ version as assumed in previous theoretical work. We derive a new expression for the wall profile that generalizes the standard Landau-Lifshitz form. Our results are found to be in very good agreement with earlier numerical work using the Monte Carlo method. Preceding theories concerning the domain wall contribution to magnetoresistance have considered the role of $D$ only through the modification of the density of states in the electronic band structure. This Brief Report reveals that the wall profile itself contains an additional $D$ dependence for the case of anisotropic exchange interactions.
\end{abstract}

\maketitle

The competition between exchange and anisotropy energies
stabilizes the ground state of ferromagnetic systems as a set of
domains with different magnetizations. \cite{bloch} In the region
between the domains the magnetization smoothly changes in a way to
continuously connect the different sectors. This configuration, known as a domain wall (DW), is relevant to understanding the transport
properties \cite{transport,tatara2,levy} and the response of such systems to
external fields. \cite{susceptibility} However, despite the fact
that much is known about the magnetic structure of bulk materials, \cite{hubert} a framework to deal with the most general situation is
not available. \cite{aharoni} In some specific situations
the theoretical limitations to solve the general problem can be
handled, assuming that the exchange interaction is
\textit{isotropic} while the anisotropy in the system, in general, 
can be described by an \textit{on-site}, also called single-spin, interaction. With
these assumptions, and using a variational approach, Landau and
Lifshitz \cite{landau} obtained the exact form of a Bloch wall.
They found the profile of the wall to be described by the expression
$\cos\theta(x)= -\tanh(x/\lambda)$, where
$\theta(x)$ is the polar angle of a classical spin vector at
position $x$ and $\lambda$ is half the effective DW width. This kind of one-dimensional distribution of the magnetic
moments is always appropriate if the surface effects can be
neglected. Under this assumption, a system in any dimension
effectively behaves as one dimensional. This simplified model is
in very good agreement with the experimental results in a wide
variety of investigated materials \cite{mikeska} where the
anisotropy mainly comes from the so-called crystalline field. On the other hand, the formation of DWs in uniaxial ferromagnets with two-spin exchange anisotropy has been much less studied. \cite{yamanaka2} Reference \onlinecite{aharoni} has mentioned in page 77 that there is no experimental evidence for the existence of this kind of anisotropy. Recently, UNiGe has shown strong indication for the presence of this anisotropic interaction. \cite{chatel} We can expect that new compounds will be discovered as part of the fast advance that experimental magnetism is currently facing in the synthesis of new materials. 

The simplest Hamiltonian describing anisotropic exchange interactions
(exchange anisotropy) is the one-dimensional (1D) $XXZ$ Heisenberg model

\begin{equation}
H_{XXZ}^{1D} = -J \sum\limits_{i}( S_{i}^{x} S_{i+1}^{x}+
S_{i}^{y} S_{i+1}^{y} + \Delta S_{i}^{z} S_{i+1}^{z})\,,
\label{1dnoonsite}
\end{equation}
where $J>0$ and $\Delta>1$ (i.e., easy-axis anisotropy). For
simplicity we will write Eq. \eqref{1dnoonsite} as

\begin{equation}
H_{XXZ}^{1D} = H_{isot}^{1D}-J \delta \sum\limits_{i} S_{i}^{z} S_{i+1}^{z}\,,
\label{1dn2}
\end{equation}
where $H_{isot}^{1D}$ is the one-dimensional isotropic Heisenberg
Hamiltonian and $\delta\equiv\Delta-1>0$. Recently, Yamanaka and
Koma \cite{yamanaka2} have used this model to describe the
formation of a DW in a system in \textit{any} dimension.
However, as will be demonstrated, the treatment of the anisotropic
exchange interactions deserves a more careful attention and, the
effective one-dimensional Hamiltonian of the problem for $D>1$ is
\textit{not} the Hamiltonian \eqref{1dnoonsite}. In fact, the
effective Hamiltonian must contain an additional on-site
anisotropy contribution whose role has been ignored for $D>1$.
This contribution has its origin in the nonlocal character of the
exchange anisotropy. On physical grounds, it contains the
information about each neighbor spin out of the chains that
support the domain structure. It is our main purpose to show the
correct derivation of the 1D-effective Hamiltonian from the
D-dimensional $XXZ$ model and obtain a closed expression to the
effective width of the wall using  both the variational
\cite{landau} and semiclassical \cite{mikeska} approaches. Our results will be compared to those obtained in Ref. \onlinecite{yamanaka2} and to earlier reported numerical Monte Carlo data. \cite{serena} In closing, we will consider previous theories concerning the DW contribution to magnetoresistance in connection to the present work.

Our starting point will be the D-dimensional $XXZ$ Heisenberg model

\begin{equation}
H_{XXZ}^{D} = H_{isot}^{D}-J \delta \sum\limits_{\langle \mathbf{a},\mathbf{b}
\rangle} S_{\mathbf{a}}^{z} S_{\mathbf{b}}^{z}\,, \label{noonsite}
\end{equation}
where the sum goes over all bounds between nearest neighbors. We
will be interested in solutions with the boundary conditions
$\lim_{a_{x}\to\mp\infty}{\mathbf S}_{\mathbf{a}}=S\,(0,0,\pm 1)$,
 a $\pi$ DW in the $x$ direction. Therefore, the
magnetization on the sample will look the same along any spin
chain in the $y$ direction [two-dimensional (2D) case] or both $y$ and $z$
directions [three-dimensional (3D) case].

The effective one-dimensional Hamiltonian can be obtained using
the fact that $S_{\mathbf{b}}^{\alpha}=S_{\mathbf{a}}^{\alpha}$
$(\alpha=x,y,z)$ for every nearest neighbor of the spin in the
$\mathbf{a}$th site, except for the ones on the $x$ axis. After
summing over the sites out of the chains in $x$ direction, the
Hamiltonian \eqref{noonsite} reduces to the 1D version

\begin{equation}
\begin{split}
H_{eff}^{1D}&= H_{isot}^{1D}-J \delta \sum\limits_{i} S_{i}^{z} S_{i+1}^{z}
-J \delta(D-1)\sum\limits_{i} \left.S_{i}^z \right.^2\\
&=H_{XXZ}^{1D}-J \delta(D-1)\sum\limits_{i} \left.S_{i}^z \right.^2\,.\label{effective}
\end{split}
\end{equation}
As can be seen, the one-dimensional effective Hamiltonian
\eqref{effective} coincides with the 1D version of the $XXZ$
model \textit{only} in the obvious case $D=1$. For $D>1$, the
exchange anisotropy leads to an additional effective on-site
contribution depending on $D$. We notice that $H_{isot}^{D}$ also
leads to a similar \textit{but} isotropic contribution depending
on $S_i^2$ which cannot affect the form of the wall. In the
usual case studied in Refs. \onlinecite{landau} and
\onlinecite{mikeska}, the effective one-dimensional Hamiltonian is
simply the 1D version of the original $D$-dimensional model
because the anisotropy is assumed to be of the on-site type. The
additional on-site term in \eqref{effective} makes the form and
the width of the wall dependent on the system dimension or on the
coordination number of the crystal structure.

The modification of the DW profile induced by the new
terms can be obtained following a  variational approach as in
Ref. \onlinecite{landau}. A crystalline on-site anisotropy given by $-J\,\delta_c\sum\limits_i\left. S_i^z \right.^2$ can be added to \eqref{effective} in order to consider a more general situation.  We assume small values of $\delta$ and $\delta_c$ so that the continuum limit makes sense. Then, Eq. \ref{effective} with the crystalline on-site term added, defines the one-dimensional variational problem given by

\begin{equation}
\int\left(\frac{J}{2}(\partial_x \mathbf{S})^2+\frac{J\delta}{2}
(\partial_x S^{z})^2-J\delta_t\left.S_{i}^z \right.^2\right)dx =
\mathrm{min}\,,\label{min}
\end{equation}
where $\delta_t\equiv\delta\,D+\delta_c$ is the total anisotropy parameter depending on $D$ and $x$ is in units of the lattice constant.

As it is shown in \eqref{min}, the exchange is anisotropic and the on-site interaction depends on the system dimension. Therefore, the DW
profile and its effective width are different from those obtained
in Ref. \onlinecite{landau}. As mentioned before, the on-site term reflects the
fact that each site effectively feels the cost of energy to have
its nearest neighbors pointing in directions away from the
anisotropy axis. As the number of nearest neighbors increases with
the dimension we can expect the effective width of the wall to
decrease for higher $D$.

Using the angle representation for the magnetic moments

\begin{equation}
\mathbf{S}(x) = S\,\left[0,\sin\theta(x),\cos\theta(x)\right],
\label{vector}
\end{equation}
where we assume $S\gg 1$, Eq. \eqref{min} becomes

\begin{equation}
\int \left[\frac{1}{2}\left(1+\delta\sin^2\theta\right)\theta'^2-\delta_t\cos^2\theta
\right]\mathrm{d}x = \mathrm{min}\,,\label{min2}
\end{equation}
which generates the equation of motion

\begin{equation}
\frac{\delta\theta'^2-2\delta_t}{2}\sin 2\theta+\left(1+\delta\sin^2\theta\right)
\theta''=0 .
\label{euler}
\end{equation}
Integration of \eqref{euler} with the boundary conditions of a
$\pi$-DW leads to the solution for the magnetic structure of the form

\begin{equation}
\begin{split}
-\sqrt{2(\delta\,D+\delta_c)}\;x\,=\,&\mathrm{arctanh}
\left[\frac{\cos\theta(x)}{\sqrt{1 + \delta\,\sin^{2}\theta(x)}}\right]\\
&+\!\sqrt{\delta}\,\arctan\left[\frac{\sqrt\delta\,\cos\theta(x)}{\sqrt{1 +
\delta\,\sin^{2}\theta(x)}}\right].
\label{eqarct}
\end{split}
\end{equation}
As can be seen, the standard Landau and Lifshitz \cite{landau} expression $\cos\theta(x)= -\tanh(\sqrt{2 \delta_c}\; x)$ can be obtained from Eq. \eqref{eqarct} for $\delta=0$.  Equation \eqref{eqarct} is thus the generalization of that expression for the case of anisotropic exchange interactions. 

The same expression for the DW profile can be obtained following
a different approach, as in Ref. \onlinecite{mikeska}. In that case, the static solution of
the spin quantum equations of motion,
\begin{equation}
\frac{dS_{\mathbf{a}}^{\alpha}}{dt}=0=\frac{1}{i\hbar} [S_{\mathbf{a}}^{\alpha},H]
,\label{motion}
\end{equation}
is considered. Here $H$ can be either the $D$-dimensional
Hamiltonian \eqref{noonsite} or the effective one-dimensional
Hamiltonian \eqref{effective}. The sum over the nearest neighbors
of the  $\mathbf{a}$th spin resulting from \eqref{motion} can be
transformed in the continuum limit as

\begin{equation}
\sum\limits_{\mathbf{b}}S_{\mathbf{b}}^{\alpha}\to\partial_{x}^{2}
S^{\alpha}(x)+ 2\,D\, S^{\alpha}(x).\label{cont}
\end{equation}
For the case of the effective one-dimensional Hamiltonian
\eqref{effective}, $D$ must be taken equal to $1$ in Eq.
\eqref{cont}. Then the classical vector representation
\eqref{vector} can be used and Eq. \eqref{euler} --and thus Eq.
\eqref{eqarct}-- is recovered. We note that according to Eq.
\eqref{vector} one finally takes $S^{x}(x)=0$. Thus, only the
evolution of $S_{\mathbf{a}}^{x}$ actually needs to be considered
in Eq. \eqref{motion}.

We can extract the effective width of the wall ($W$) from the behavior of
Eq. \eqref{eqarct} near the origin. For $x\to 0$ we get
$\sqrt{1+\delta}\,\cos\theta(x) = -\sqrt{2(\delta\,D+\delta_c)}\;x$ and, therefore,

\begin{equation}
W=2\sqrt{\frac{1+\delta}{2\delta_t}}=2\sqrt{\frac{1+\delta}{2(\delta\,D+\delta_c)}}\,.
\label{eqest}
\end{equation}
As expected, an increase in the dimension results in a narrower
DW. This is in agreement with the fact that the local
field produced by the spins out of a given chain increases the
energy cost of the spin twist. As a consequence, the system
effectively gets less spins out of the $z$ axis and thus the
effective width of the wall becomes smaller. The specific behavior $\propto1/\sqrt{\delta_t}$ agrees with the limit case $\delta\to0$, in which
the anisotropic exchange term $(J\delta/2)(\partial_x S^{z})^2$ in
Eq. \eqref{min} can be neglected. In that case, one gets an
isotropic exchange parameter $J/2$ and an on-site anisotropy with
the effective parameter $J\,\delta_t$. The ratio of these two
parameters determines \cite{landau} the effective width of the
wall. Slightly away from that limit, Eq. \eqref{eqest} shows an
additional exchange-type dependence on $\delta$ (i.e., $\delta$ is
in the numerator) coming from the total exchange parameter
$1+\delta=\Delta$ of the $z$ components in the $XXZ$ model.

When $\delta=0$, Eq. \eqref{eqest} gives $2\frac{1}{\sqrt{2\delta_c}}$, which has no dependence on $D$ and is the effective width of the wall  in the well-known case for which the
anisotropy in the original model is of the on-site
type. \cite{landau,mikeska} As can be seen from \eqref{eqest}, the effective width has less sensitivity to variations in the parameter $\delta$ than in $\delta_c$. This is a consequence of the combined role of $\delta$ as exchange and
anisotropy parameter at the same time, as can be seen from the original model \eqref{noonsite}.

Now we are going to compare our results for $D=1$ with those obtained in
Ref. \onlinecite{yamanaka2}, where $\delta_c=0$ was considered. Following a pure quantum method in treating
the $XXZ$ model \eqref{1dnoonsite}, Yamanaka and
Koma \cite{yamanaka2} rederived the
Landau and Lifshitz expression $\cos\theta(x)= -\tanh(x/\lambda)$
with an effective wall width given by

\begin{equation}
2\lambda =-2\frac{1}{\ln\,{q}},
\label{estq}
\end{equation}
where $0<q<1$ and $q +1/q =2\Delta$.

As discussed above, we can get such an expression for the domain
wall profile, which corresponds to the case of isotropic exchange
interactions and on-site anisotropy, \textit{only} in the limit
case $\delta\to0$. Indeed, it is not hard to see that for $D=1$
and $\delta\to0$, Eq. \eqref{eqarct} --with $\delta_c=0$-- reduces to that expression
with $\lambda$ given by $1/\sqrt{2\delta}$.

At first sight the DW width \eqref{estq} looks quite
different from our result given by Eq. \eqref{eqest}, with $\delta_c=0$, in the $D=1$
case. However, as can be seen they also coincide for small
values of $\delta$ ($\Delta\approx1$). One first notices that
since $\Delta>1$, then $q=\Delta-\sqrt{\Delta²-1}$. We can expand
Eqs. \eqref{eqest} and \eqref{estq} in powers of $\delta$ and
obtain for both cases
$2/\sqrt{2\delta}+\mathcal{O}(\delta^{1/2})$. \cite{note1} Thus, up
to leading order in $1/\delta$, both the expression \eqref{eqest}, with $\delta_c=0$, and
the one reported in Ref. \onlinecite{yamanaka2} are equal to
$2\frac{1}{\sqrt{2\delta}}$. It is interesting to note that the
Eq. \eqref{eqarct} for the form of the wall and \eqref{eqest} for
the effective DW width have been derived for large values
of spin $S$. However, both of them reduce, for $D=1$ and small
values of the parameter $\delta$, to the results of Ref.
\onlinecite{yamanaka2} for spin $S=1/2$.

The anisotropic Heisenberg model has been studied before using the Monte Carlo method. \cite{mc} Recently, Serena and Costa-Krämer \cite{serena} have made use of this technique to describe a $\pi$ DW in the $XXZ$ model. In order to compare our results with theirs, we first make the equivalence between our parameters and those used in Ref. \onlinecite{serena}. Hereafter, to avoid confusion with $\Delta$ in the present work, we identify the anisotropy parameter of Ref. \onlinecite{serena} by $\Delta'$. One can see that they are related by $\Delta=1/(1-\Delta')$. Therefore $\delta=\Delta-1=\Delta'/(1-\Delta')$ and, from Eq. \eqref{eqest}, with $\delta_c=0$, we get

\renewcommand{\figurename}{Fig.}

\newcommand{\captionfonts}{\small}

\begin{figure}
\centering
\includegraphics[width=\linewidth]{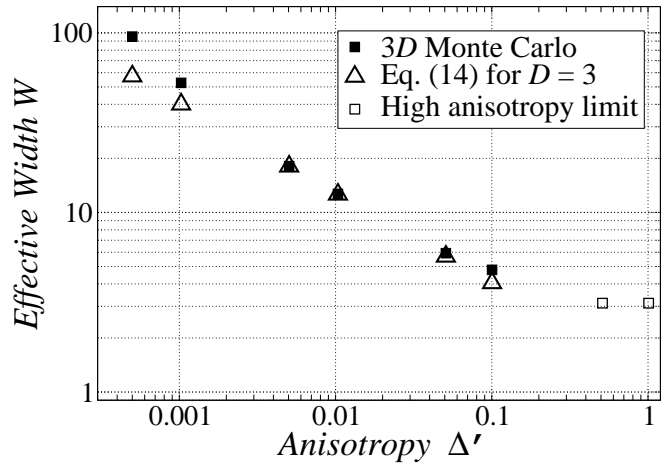}
\caption{The effective DW width, as obtained from Eq. \eqref{eqest1} for $D=3$ (open triangles), is compared to Monte Carlo data (filled squares) from Ref. \onlinecite{serena}. Open squares correspond to the high anisotropy limit for which Eq. \eqref{eqest1} is not applicable. Please see text for further details.} 
\label{figdemc} 
\end{figure}

\begin{equation}
W = \pi\frac{1}{\sqrt{2\,\Delta'\,D}}\,.
\label{eqest1}
\end{equation}
Here, the factor $2$ of Eq. \eqref{eqest} has been changed to $\pi$ to follow the definition used in Ref. \onlinecite{serena} for the effective DW width. Equation \eqref{eqest1} gives $W$ in a simpler manner than the one given by Eq. \eqref{eqest} for $\delta_c=0$. Additionally, it looks closer to the result $\pi\frac{1}{\sqrt{2\delta_c}}$ of the usual case of isotropic exchange interactions and on-site anisotropy. Equation \eqref{eqest1} confirms the power-law $W\propto\Delta'^{-1/2}$ that has been suggested in Ref. \onlinecite{serena}. It also elucidates the fact that the proportionality factor of this behavior is determined by the dimensionality $D$ of the system.    

Figure \ref{figdemc} compares the effective DW width, as obtained from Eq. \eqref{eqest1}, to the numerical data from Ref. \onlinecite{serena} for the  case $D=3$.  Here some remarks are in order. Equation \eqref{eqest1} has been derived in the continuum limit, i.e., large values of $W$ and thus small anisotropy values have been assumed. We need $\Delta'\ll1/(2\,D)$ in Eq. \eqref{eqest1} to get $W\gg1$. Therefore, for $D=3$, we can expect Eq. \eqref{eqest1} to give good results for $\Delta'\ll1/6\approx0.2$. This agrees with the fact observed by Serena and Costa-Krämer that for anisotropy values larger than $\Delta'=0.5$, the magnetization suddenly changes in one lattice constant. \cite{serena} The value $W=\pi$ has been defined in Ref. \onlinecite{serena} for this high anisotropy limit. Open squares in Fig. \ref{figdemc} correspond to this case and are not considered in Eq. \eqref{eqest1}. An unexpected departure from Eq. \eqref{eqest1} is observed for Monte Carlo (MC) data with very small anisotropy values. According to Ref.  \onlinecite{serena}, this may be associated to numerical uncertainties for such very low values of anisotropy ($\Delta'<0.001$). With these comments in mind, we conclude that our results are in very good agreement with the numerical data of Serena and Costa-Krämer. 

Previous theories have considered the role of a DW in the transport properties of ferromagnetic metals. The effective width $W$, when compared to the Fermi wavelength $\lambda_F$, determines how strong the effect of the DW is in increasing \cite{transport,levy} or decreasing \cite{tatara2} the electrical resistance. Recently, Sil and Entel \cite{sil} have considered the modification of the electronic band structure by the DW and shown that there can be an increase or decrease of the resistance depending on the dimensionality, position of the Fermi energy and strength of the coupling between conduction electrons and DW. The role of $D$ has been considered through the modification of the density of states. One can see that these theories have considered no dependence of the effective DW width on the dimensionality $D$ of the system. Here we have seen that this is valid only for the usual case of isotropic exchange interactions and on-site anisotropy. This Brief Report reveals that the wall profile itself contains an additional $D$ dependence for the case of anisotropic exchange interactions. In this case Eq. \eqref{eqest} will define such a $D$ dependence of $W$, which is the relevant DW parameter in the electronic transport. 

To summarize, we have followed both the variational and
semiclassical approaches to describe a $\pi$ DW in the
$XXZ$ model in a $D$-dimensional system. It was shown that the
effective one-dimensional Hamiltonian contains an on-site
anisotropy contribution which has been previously ignored for
$D>1$. The standard Landau-Lifshitz DW profile for isotropic exchange interactions has been generalized. Increasing the dimension of the system decreases the width of
the wall. This case is thus essentially different from the well known one
for which the anisotropy in the original model is only of the on-site
type. Very good agreement with earlier numerical data was found. This work adds a new source of $D$ dependence for measurable physical
properties like the magnetoresistance.

\begin{acknowledgments}
A.O.G.R. would like to acknowledge E. Miranda and G. G. Cabrera for very useful discussions and Coordenação de Aperfeiçoamento de Pessoal de Nível Superior (CAPES) for financial support. The same author gratefully acknowledges help from R. Lora and P. A. Serena concerning Ref. \onlinecite{serena}. A.V.F. and A.O.C. wish to thank Fundação de Amparo à Pesquisa do Estado de São Paulo (FAPESP), whereas A.O.C. also acknowledges partial support from Conselho Nacional de Desenvolvimento Científico e Tecnológico (CNPq).
\end{acknowledgments}

\bibliography{DWbib}

\begin{thebibliography}{15}
\bibitem[{not()}]{note1}
\bibinfo{note}{In a less precise way, we have that since $\Delta\approx1$ then
  $\ln{q}=\ln{\Delta}+\ln{\left(1-\frac{\sqrt{\Delta² -
  1}}{\Delta}\right)}\approx-\frac{\sqrt{\Delta² -
  1}}{\Delta}\approx-\sqrt{\Delta² - 1}=-\sqrt{\Delta+1}\sqrt{\Delta -
  1}\approx-\sqrt{2}\sqrt{\Delta - 1}=-\sqrt{2\delta}$}.
\end{thebibliography}

\end{document}